%% file: main.tex
\PassOptionsToPackage{table}{xcolor}

\documentclass[letterpaper]{article} % DO NOT CHANGE THIS

\ifdefined\aaaianonymous
    \usepackage[submission]{aaai2026}  % Anonymous submission version
\else
    \usepackage{aaai2026}              % Camera-ready version
\fi

\usepackage{times}  % DO NOT CHANGE THIS
\usepackage{helvet}  % DO NOT CHANGE THIS
\usepackage{courier}  % DO NOT CHANGE THIS
\usepackage[hyphens]{url}  % DO NOT CHANGE THIS
\usepackage{graphicx} % DO NOT CHANGE THIS
\usepackage{natbib}  % DO NOT CHANGE THIS AND DO NOT ADD ANY OPTIONS TO IT
\usepackage{caption} % DO NOT CHANGE THIS AND DO NOT ADD ANY OPTIONS TO IT
\usepackage{pifont}
\usepackage{multirow}
\usepackage{enumitem}
\usepackage{algorithm}
\usepackage{algorithmic}
\usepackage{amsmath,amssymb}
\usepackage{booktabs} 
\usepackage{tabularx}
\usepackage{textcomp}
\usepackage{newfloat}
\usepackage{listings}
\usepackage{xcolor} 
\usepackage{colortbl} 
\definecolor{rowgray}{HTML}{EFEFEF}
\DeclareCaptionStyle{ruled}{labelfont=normalfont,labelsep=colon,strut=off} % DO NOT CHANGE THIS
\floatstyle{ruled}
\newfloat{listing}{tb}{lst}{}
\floatname{listing}{Listing}

\ifdefined\aaaianonymous
    \title{SafeSieve: From Heuristics to Experience in Progressive Pruning for LLM-based Multi-Agent Communication}
\else
    \title{SafeSieve: From Heuristics to Experience in Progressive Pruning for LLM-based Multi-Agent Communication}
\fi

\author{
    Ruijia Zhang\textsuperscript{\rm 1},
    Xinyan Zhao\textsuperscript{\rm 1},
    Ruixiang Wang\textsuperscript{\rm 1},
    Sigen Chen\textsuperscript{\rm 1},
    Guibin Zhang\textsuperscript{\rm 1},
    An Zhang\textsuperscript{2,$\dagger$},
    Kun Wang\textsuperscript{\rm 3},
    Qingsong Wen\textsuperscript{\rm 4}
}
\affiliations{
    \textsuperscript{\rm 1} National University of Singapore, Singapore \\
    \textsuperscript{\rm 2} School of Public Policy and Administration, Chongqing University, China \\
    \textsuperscript{\rm 3} Nanyang Technological University, Singapore \\
    \textsuperscript{\rm 4} AI Research Institute, Squirrel Ai Learning, China \\
    e0536882@u.nus.edu, zhao\_xinyan@u.nus.edu, ruixiangw@u.nus.edu, e1538182@u.nus.edu, \\
    guibinz@outlook.com, 20210101013@cqu.edu.cn, wk520529@mail.ustc.edu.cn, qingsongedu@gmail.com \\
}

\usepackage{bibentry}
\DeclareUnicodeCharacter{202F}{\,}
\def\nocopyright{\gdef\copyright@on{}}

\begin{document}

\maketitle

\begin{abstract}
LLM-based multi-agent systems exhibit strong collaborative capabilities but often suffer from redundant communication and excessive token overhead. Existing methods typically enhance efficiency through pretrained GNNs or greedy algorithms, but often isolate pre- and post-task optimization, lacking a unified strategy. To this end, we present SafeSieve, a progressive and adaptive multi-agent pruning algorithm that dynamically refines the inter-agent communication through a novel dual-mechanism. SafeSieve integrates initial LLM-based semantic evaluation with accumulated performance feedback, enabling a smooth transition from heuristic initialization to experience-driven refinement. Unlike existing greedy Top-k pruning methods, SafeSieve employs 0-extension clustering to preserve structurally coherent agent groups while eliminating ineffective links. Experiments across benchmarks (SVAMP, HumanEval, etc.) showcase that SafeSieve achieves 94.01\% average accuracy while reducing token usage by 12.4\%-27.8\%. Results further demonstrate robustness under prompt injection attacks (1.23\% average accuracy drop). In heterogeneous settings, SafeSieve reduces deployment costs by 13.3\% while maintaining performance. These results establish SafeSieve as an efficient, GPU-free, and scalable framework for practical multi-agent systems. Our code can be found below.
\end{abstract}

% Links section - only shown in camera-ready version
% \ifdefined\aaaianonymous
\begin{links}
    \link{Code}{https://github.com/csgen/SafeSieve}
\end{links}

\def\addcontentsline#1#2#3{}
% gf: PRINT COPYRIGHT NOTICE
\def\copyright@year{\number\year}
\def\copyright@text{Copyright \copyright\space \copyright@year,
Association for the Advancement of Artificial Intelligence (www.aaai.org).
All rights reserved.}
\def\copyrighttext#1{\gdef\copyright@on{T}\gdef\copyright@text{#1}}
\def\copyrightyear#1{\gdef\copyright@on{T}\gdef\copyright@year{#1}}

% % Version-specific content
% \ifdefined\aaaianonymous
% \section{Preparing an Anonymous Submission}

% This document details the formatting requirements for anonymous submissions. The requirements are the same as for camera ready papers but with a few notable differences:

% \begin{itemize}
%     \item Anonymous submissions must not include the author names and affiliations. Write ``Anonymous Submission'' as the ``sole author'' and leave the affiliations empty.
%     \item The PDF document's metadata should be cleared with a metadata-cleaning tool before submitting it. This is to prevent leaked information from revealing your identity.
%     \item References must be anonymized whenever the reader can infer that they are to the authors' previous work.
%     \item AAAI's copyright notice should not be included as a footer in the first page.
%     \item Only the PDF version is required at this stage. No source versions will be requested, nor any copyright transfer form.
% \end{itemize}

% You can remove the copyright notice and ensure that your names aren't shown by including \texttt{submission} option when loading the \texttt{aaai2026} package:

% \begin{quote}\begin{scriptsize}\begin{verbatim}
% \documentclass[letterpaper]{article}
% \usepackage[submission]{aaai2026}
% \end{verbatim}\end{scriptsize}\end{quote}

% The remainder of this document are the original camera-ready instructions. Any contradiction of the above points ought to be ignored while preparing anonymous submissions.

% \section{Introduction}
% \else
\section{Introduction}
%\fi

Large language model~(LLM) based multi-agent systems~(MAS)
have demonstrated impressive collaborative problem-solving capabilities~\cite{wang2025comprehensive,chang2024survey}, fueling frameworks such as AutoGen and ChatDev for real-world applications \cite{AutoGen2023, chatdev}. Nevertheless, the dense, round-robin conversations among agents often incur substantial token overhead and communication redundancy, which not only elevates inference cost but also dilutes attention over key information, leading to potential accuracy degradation \cite{Liu2023LostMid}. Longer context windows further enlarge the attack surface for prompt-injection \cite{anil2024manyshot}. Consequently, recent studies have begun to sparsify MAS communication topologies to improve both efficiency
and robustness \cite{zhuge2024gptswarm, Zhang2024GDesigner, Zhang2024AgentPrune}.

\begin{figure}[t]
\centering \includegraphics[width=\columnwidth]{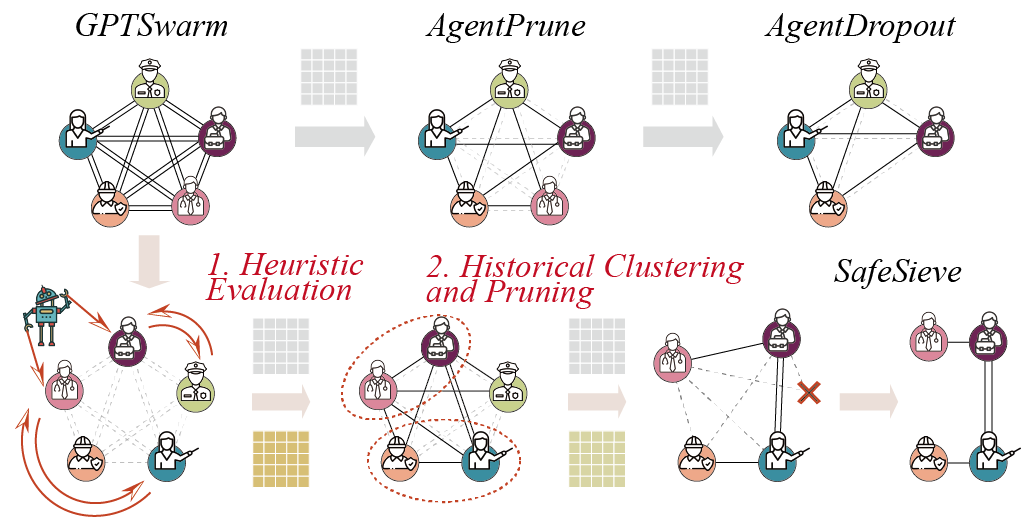}
\caption{Comparison of \textbf{SafeSieve} with GPTSwarm, AgentPrune, and AgentDropout. It illustrates the evolutionary trajectory of post-pruning MAS, highlighting SafeSieve's novel contribution as a unified design that bridges early-stage heuristics and feedback-driven refinement.}
\label{comparison}
\end{figure}

Among communication sparsification strategies, one prevalent line of work directly constructs compact graph topologies prior to execution, such as G-Designer and GPTSwarm \cite{Zhang2024GDesigner,zhuge2024gptswarm}. These methods improve communication efficiency at initialization but exhibit limited generalizability and cannot adapt to run-time dynamics. A more recent and adaptive paradigm is on-the-fly pruning, where the communication graph is dynamically adjusted based on task performance feedback. Representative works like AgentPrune and AgentDropout \cite{Zhang2024AgentPrune,Wang2025AgentDropout} start from a basic or fully-connected topology and iteratively prune edges during execution. These approaches require no pretraining and offer strong task adaptability with minimal deployment burden. However, most of them still rely on greedy \emph{top-$k$} pruning strategies, which may mistakenly remove critical communication paths, reducing system robustness. To date, \textit{no method} has yet unified both heuristic-driven early filtering and performance-aware dynamic adaptation, forming a full-spectrum optimization pipeline.

Motivated by the similarity between MAS collaboration and human team
organization \cite{guo2024embodied_llm_teams}, we propose \textbf{SafeSieve}, a progressive and adaptive pruning algorithm. SafeSieve introduces a two-stage edge scoring scheme that \ding{192}~uses LLM-based semantic compatibility to offer heuristic guidance at startup and \ding{193}~gradually shifts weight to accumulated contribution during execution, emulating the “plan-then-adjust” paradigm of human teamwork, as shown in Figure~\ref{comparison}. Instead of pruning edges individually, SafeSieve employs a \emph{0-extension} based clustering
mechanism~\cite{fakcharoenphol2003improved0extension}, preserving structurally coherent agent groups while eliminating ineffective links. This design avoids the local sub-optimality of greedy top-$k$ pruning and retains inter-agent complementarity.

% Check the accuracy number.

Comprehensive experiments on six benchmarks (including GSM8K, SVAMP, HumanEval, AQuA, MMLU, MATH-500) showcase that SafeSieve reduces token usage by 12.4–27.8\% and boosts accuracy by up to 2.22\%, consistently outperforming prior sparsification methods. It further remains resilient under prompt-injection, suffering only a 1.23-1.94\% accuracy drop, and supports heterogeneous collaboration where large LLMs guide smaller ones, thus expanding the real-world deployment space.

Our main contributions are threefold:
\begin{itemize}[leftmargin=*]
\item \textbf{Unified Framework.} We categorize MAS communication optimization into \textit{pre-design} and \textit{post-prune} paradigms, and propose SafeSieve—the first post-pruning framework that integrates LLM-based semantic evaluation, cumulative historical feedback, and 0-extension clustering to achieve progressive graph sparsification while preserving agent complementarity.
\item \textbf{Efficiency with Robustness.} Extensive experiments across six benchmarks demonstrate that SafeSieve reduces token consumption by 12.4\%––27.8\% compared to peer methods while maintaining or improving accuracy by up to 2.22\%. Uniquely among post-prune optimizers, SafeSieve exhibits inherent adversarial resilience, detecting and mitigating malicious agents with 1.23\% degradation.
\item \textbf{Heterogeneous Deployment.} We pioneer heterogeneous multi-agent evaluation by systematically analyzing cross-model collaboration with real-time cost tracking, revealing that SafeSieve's clustering mechanism effectively leverages model diversity to reduce deployment costs by up to 13.3\% in production settings.
\end{itemize}

\section{Related Work and Preliminary}

\paragraph{Communication Efficiency in MAS.}
Recent research in LLM-based MAS has explored two primary paradigms for optimizing communication efficiency: \textit{pre-design} approaches that construct optimized communication structures before the task, and \textit{post-prune} methods that start with various basic topologies and iteratively remove redundant connections. They reflect the trade-off between upfront design complexity and runtime adaptability in MAS.

Pre-design methods, including GPTSwarm~\cite{zhuge2024gptswarm}, G-Designer~\cite{Zhang2024GDesigner}, AnyMAC~\cite{wang2025anymac}, EvoMAC~\cite{hu2024selfevolving}, and DyLAN~\cite{liu2024dynamicllm}, focus on directly generating efficient communication topologies—whether through GNN-based graph construction, autoregressive agent selection, evolutionary adaptation, or two-stage team formation. 

Post-pruning approaches begin with dense communication topology and progressively sparsify them: AgentPrune~\cite{Zhang2024AgentPrune} introduces dynamic edge pruning via one-hot mask matrices, AgentDropout~\cite{Wang2025AgentDropout} extends this to node-level pruning, Adaptive Graph Pruning~\cite{li2025adaptivegraph} jointly learns node selection and edge connectivity via end-to-end GNN training, and Adaptive Prompt Pruning~\cite{dong2024promptprompted} reduces per-agent prompt lengths to save token. 

\paragraph{Graph Clustering and 0-extension.}
Recent studies have shown that LLMs exhibit human-like collaborative patterns, where agents with similar or complementary capabilities naturally form effective working groups~\cite{wang2023unleashing,hong2023metagpt}. Inspired by this observation, we propose that clustering-based pruning offers a more principled approach than direct edge removal in MAS collaboration. 

While various clustering methods exist—including spectral clustering~\cite{schaeffer2007graph}, hierarchical clustering~\cite{xue2024comprehensive}, and density-based approaches~\cite{birant2007stdbscan}—the $k$-terminal \textit{0-extension problem}~\cite{calinescu2003approximation,fakcharoenphol2003improved0extension} presents unique advantages for our setting. 0-extension has been successfully applied in computing system for its computational efficiency ($O(n\log n)$ complexity), simple deployment, and strong theoretical guarantees in preserving graph connectivity~\cite{englert2014vertex,chen2011clustering}. By formulating agent clustering as a 0-extension problem, we replace aggressive top-$k$ pruning with a connectivity-aware approach that maintains critical communication paths between agent communities while achieving similar sparsification rates.

\paragraph{MAS as a Communication Graph.}
Building upon the graph-based paradigm for multi-agent systems, GPTSwarm~\cite{zhuge2024gptswarm} first formalized multi-agent orchestration as a differentiable computational graph. At each communication round $t$, the interaction among $N$ agents is represented as a directed communication graph $\mathcal{G}_t=(\mathcal{V}, \mathcal{E}_t)$, where $\mathcal{V}$ denotes the set of agents and each edge $e_{ij} \in \mathcal{E}_t$ represents a message from agent $i$ to agent $j$. While GPTSwarm generates static graph structures during inference, AgentPrune~\cite{Zhang2024AgentPrune} introduces dynamic sparsification through a mask matrix $\mathbf{M}^{(t)} \in \{0,1\}^{N \times N}$, where each entry controls edge activation:
\begin{equation}
\tilde{\mathcal{E}}_t = \{e_{ij} \in \mathcal{E}_t : M_{ij}^{(t)} = 1\}
\end{equation}
This mask effectively defines a candidate set of communication links from which the actual runtime communication graph is sampled. AgentDropout~\cite{Wang2025AgentDropout} extends this framework by incorporating node-level pruning and real-time feedback mechanisms, enabling more aggressive sparsification across communication rounds.

\begin{figure*}[t]
\centering
\includegraphics[width=\textwidth]{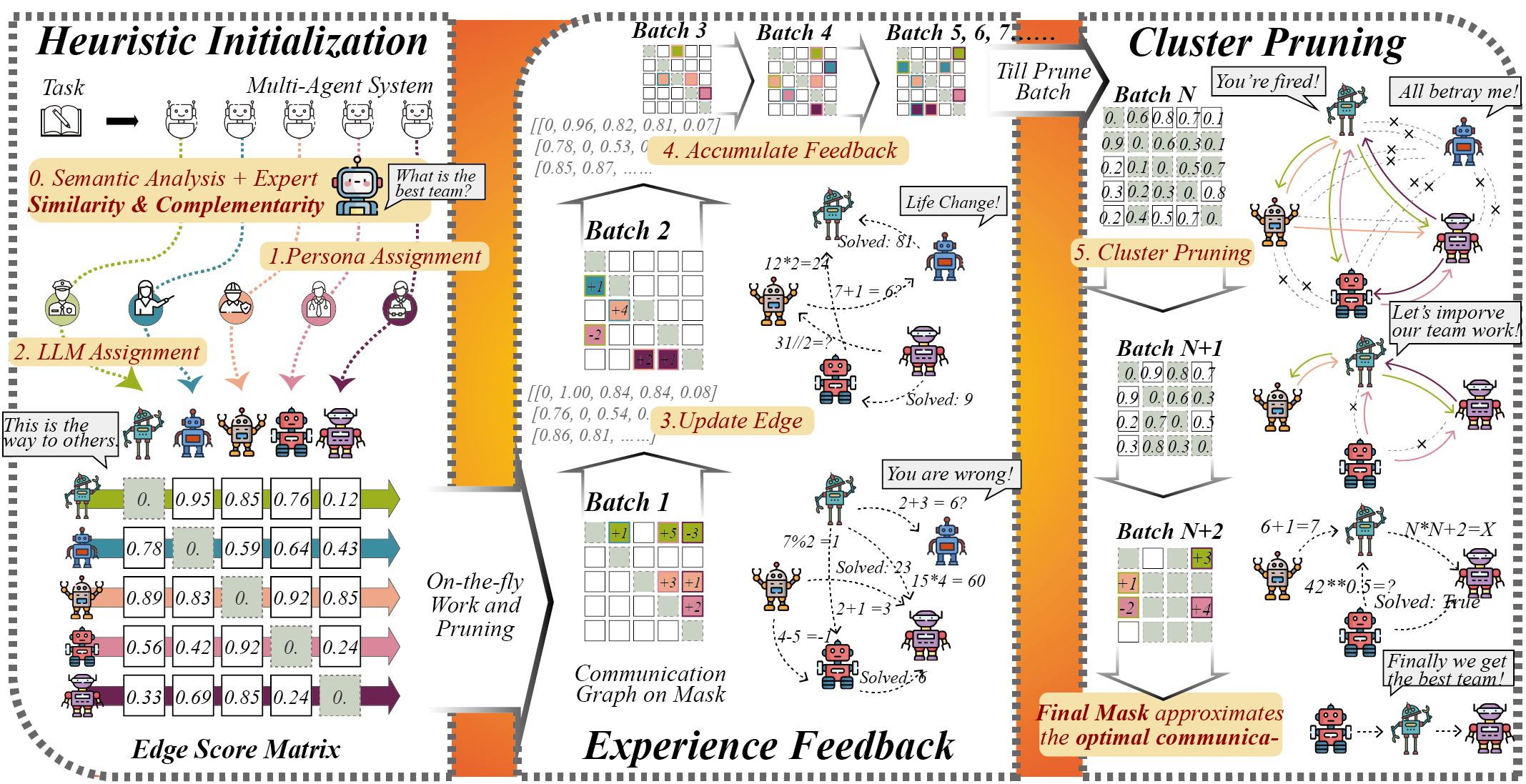} 
\caption{\textbf{\textit{SafeSieve Pipeline}}. The process begins by constructing a complete communication graph based on semantic relevance among agent roles. During task execution, edge importance is updated based on reasoning success, enabling adaptive pruning via 0-extension clustering. The final communication structure reflects a task-aware, resource-efficient collaboration topology.}
\label{pipeline}
\end{figure*}

\section{SafeSieve: Integrated Pruning Strategy}

SafeSieve progressively prunes communication graphs in MAS by assessing inter-agent link quality through a dynamic edge scoring matrix $E \in \mathbb{R}^{n \times n}$. We combine semantic compatibility—obtained from expert LLM assessments—with historical complementarity based on interaction outcomes. This dual mechanism enables adaptive edge pruning based on \textbf{time-varying} thresholds, while isolated nodes are naturally removed. 0-extension clustering is integrated to preserve coherent community structures and information flow. Figure~\ref{pipeline} illustrates the whole pruning process.

\subsection{Semantic Heuristic Initialization}

While semantic similarity facilitates basic cooperation~\cite{deng2025semantic_multi_agent_comm}, functional complementarity plays a more critical role in complex multi-hop reasoning tasks~\cite{zhang2024chain_of_agents,zhang2025multi_agent_collaboration_survey}. SafeSieve initializes the communication edge score between agents $i$ and $j$ by combining embedding-based similarity with expert-assessed compatibility:

\begin{equation}
S_{ij}^{\text{compat}} = \gamma \cdot \frac{\mathbf{e}_i \cdot \mathbf{e}_j}{||\mathbf{e}_i|| \cdot ||\mathbf{e}_j||} + (1 - \gamma) \cdot \mathcal{Q}(S_{ij}^{\text{expert}})
\end{equation}

where $\mathbf{e}_i, \mathbf{e}_j \in \mathbb{R}^d$ are pre-trained role embeddings that capture agent capabilities, $S_{ij}^{\text{expert}} \in [0,1]$ represents the functional compatibility score assessed by an expert LLM, and $\mathcal{Q}(\cdot)$ denotes a 5-level quantization function that discretizes continuous scores into categorical levels. The balance parameter $\gamma \in [0,1]$ controls the relative importance of semantic similarity versus expert-assessed complementarity.

\subsection{Progressive Pruning with Historical Feedback}

To capture the dynamic nature of agent cooperation, SafeSieve progressively shifts from static semantic initialization to experience-based scoring through a unified temporal mechanism~\cite{zhang2021multiagent_deep_rl}. 

\paragraph{Historical Complementarity.} The system tracks each edge's contribution to successful task completion over time. The historical complementarity score quantifies the accumulated value of edge $(i,j)$ relative to all edges in the graph:

\begin{equation}
C_{ij}^{\text{hist}}(t) = \frac{\sum_{\tau=1}^{t} \mathbf{1}_{ij}^{\text{correct}}(\tau)}{\sum_{(k,l) \in E_t} \sum_{\tau=1}^{t} \mathbf{1}_{kl}^{\text{correct}}(\tau) + n^2\varepsilon}
\end{equation}

Here, $\mathbf{1}_{ij}^{\text{correct}}(\tau) \in \{0,1\}$ indicates whether edge $(i,j)$ contributed to a correct answer at time step $\tau$, $E_t$ is the set of active edges at time $t$, and $n$ is the total number of agents. 

Throughout SafeSieve, we use $\varepsilon > 0$ as a small constant to ensure numerical stability and prevent division by zero. The normalization term $n^2\varepsilon$ also accounts for the maximum possible number of edges in the complete graph.

\paragraph{Integrated Edge Scoring.} The overall edge score dynamically combines semantic initialization with accumulated historical feedback, gradually emphasizing learned patterns over static heuristics with chronological weights:

\begin{equation}
\begin{split}
E_{ij}(t) =\; & \left(1 - \frac{t}{T}\right) \cdot \alpha_0 \cdot S_{ij}^{\text{compat}} \\
            +\; & \left[\beta_0 + (\beta_{\text{max}} - \beta_0) \cdot \frac{t}{T} \right] \cdot C_{ij}^{\text{hist}}(t)
\end{split}
\end{equation}

where $\alpha_0 > 0$ is the initial weight for semantic compatibility, $\beta_0 \geq 0$ and $\beta_{\text{max}} > \beta_0$ define the range of historical contribution weights, $t$ is the current time step, and $T$ is the total number of time steps. This formulation ensures a smooth transition from heuristics to experience evaluation.

\subsection{0-Extension Clustering for Pruning Decisions}

SafeSieve employs a principled clustering approach based on the 0-extension framework~\cite{fakcharoenphol2003improved0extension} to make globally-informed pruning decisions. This method provides theoretical approximation guarantees while maintaining computational and communicative efficiency.

\paragraph{Dynamic Threshold.}
The pruning threshold adapts over time to balance exploration and exploitation, starting conservative and growing aggressive as the system evolves:

\begin{equation}
\theta(t) = \theta_0 + (\theta_{\max} - \theta_0) \cdot \left[1 - \exp^{-k \cdot \max\left(\frac{t}{T}, 0\right)}\right]
\end{equation}

where $\theta_0 \geq 0$ and $\theta_{\max} > \theta_0$ are the initial and maximum threshold values, and $k > 0$ is the growth rate parameter controlling how quickly the threshold increases.

\paragraph{Terminal Selection and Cluster Assignment.}
The clustering process begins by selecting a subset of terminal agents that serve as cluster centers. Terminals are selected to maximize overall connectivity transferred from edge scores:

\begin{equation}
T = \arg\max_{S \subseteq V, |S|=|T|} \sum_{v \in S} \sum_{u \in V} \frac{1}{(E_{vu}(t) + \varepsilon)^{-1}}
\end{equation}

where $V$ is the set of all agents. The number of terminals $|T|$ is determined adaptively based on the graph size: $|T| = \max(2, \min(\sqrt{n}, \lfloor n/3 \rfloor))$, ensuring at least two clusters while avoiding over-fragmentation. Each agent is then assigned to a terminal by solving the 0-extension problem:

\begin{equation}
f^* = \arg\min_{f: V \rightarrow T} \sum_{(i,j) \in E} (E_{ij}(t) + \varepsilon)^{-1} \cdot \mathbf{1}\{f(i) \ne f(j)\}
\end{equation}

This optimization finds the cluster assignment $f^*$ that minimizes the total distance of edges crossing cluster boundaries, where distance is inversely proportional to edge score.

\paragraph{Structured Edge Pruning.}
Pruning occurs at time steps $t$ when both conditions are met: $t \geq B_{\text{start}}$ (warm-up period completed) and $\mathcal{R}(t) < R_{\max}$ (current pruning rate below maximum). The pruning set is constructed hierarchically to meet the target pruning rate $r \in (0,1)$:

\begin{equation}
|\mathcal{E}_{\text{prune}}(t)| = r \cdot |\mathcal{E}_{\text{active}}^{(t)}| = |\mathcal{E}_{\text{rule}}(t) \cup \mathcal{E}_{\text{budget}}(t)|
\end{equation}

The rule-based set $\mathcal{E}_{\text{rule}}(t)$ consists of edges that both cross cluster boundaries and fall below the threshold, specifically those satisfying the condition $(i,j)$ such that $f^(i) \ne f^(j)$, $i,j \notin T$, and $\hat{E}_{ij}(t) < \theta(t)$. If this set is less than the pruning budget, $\mathcal{E}_{\text{budget}}(t)$ supplements it by adding the lowest-scoring remaining edges until the target is reached.

\paragraph{Mask and Node Update.}
The communication mask matrix $\mathbf{M} \in \{0,1\}^{n \times n}$ is updated to reflect pruned edges:

\begin{equation}
\mathbf{M}_{ij}^{(t+1)} = \mathbf{M}_{ij}^{(t)} \cdot \mathbf{1}\{(i,j) \notin \mathcal{E}_{\text{prune}}(t)\}
\end{equation}

After edge pruning, the node set is updated to remove isolated nodes only if a minimum viable graph is maintained:

\begin{equation}
V_{t+1} = \begin{cases}
V_t \setminus \mathcal{V}_{\text{iso}}^{(t+1)} & \text{if } |V_t \setminus \mathcal{V}_{\text{iso}}^{(t+1)}| > 2 \\
V_t & \text{otherwise}
\end{cases}
\end{equation}

where $\mathcal{V}_{\text{iso}}^{(t+1)} = \{v \in V : \sum_{u \in V} \mathbf{M}_{vu}^{(t+1)} = 0\}$ denotes the set of isolated nodes after pruning.

\paragraph{Post-Pruning Regularization.}
To adapt to the changed graph structure, edge scores are normalized and historical weights are adjusted after each pruning step:

\begin{equation}
\hat{E}_{ij}(t) = \frac{E_{ij}(t) - \mu_t}{\sigma_t + \varepsilon}, \quad \hat{\beta}(t) = \beta(t) \cdot \frac{\Delta_{\text{before}}}{\Delta_{\text{after}} + \varepsilon}
\end{equation}

where $\mu_t$ and $\sigma_t$ are the mean and standard deviation of edge scores before pruning, and $\Delta_{\text{before}}$ and $\Delta_{\text{after}}$ represent the score range before and after pruning respectively. This regularization ensures that the scoring mechanism remains calibrated despite the evolving graph topology.

Further implementation details, including pseudocode and case study, are provided in Supplementary Material.

\input{tab/tab_agentperformance}
\input{tab/ablation}

\section{Experiments}

\subsection{Experiment Setup}

\paragraph{\textbf{Models \& Benchmarks.}}
In our main experiments, we adopt Deepseek-V3 (671B)~\cite{Liu2024Deepseek} as the primary backbone model. For smaller-model ablation, we use GPT-4o-mini. In heterogeneous settings, we additionally incorporate LLaMA3-8B~\cite{llama3modelcard}, Qwen2.5-72B~\cite{qwen2.5}, and Kimi-K2~\cite{Moonshot2024Kimi} to validate cross-model communication robustness. Build upon these, we evaluate general and mathematical reasoning using six standard benchmarks: MMLU~\cite{hendryckstest2021}, GSM8K~\cite{cobbe2021gsm8k}, SVAMP~\cite{patel-etal-2021-nlp}, HumanEval~\cite{chen2021evaluating}, AQuA~\cite{ling2017program} and MATH-500~\cite{lightman2023lets}.

\paragraph{\textbf{Baselines.}}
In the main experiments, we compare SafeSieve with prompting-based strategies including Vanilla (direct reasoning) and Chain-of-Thought (CoT)~\cite{wei2023chainofthoughtpromptingelicitsreasoning}(referred to as \textit{single}), as well as collaborative frameworks such as GPT-Swarm~\cite{zhuge2024gptswarm} and G-Designer~\cite{Zhang2024GDesigner}, which enhance agent outputs prior to inference (referred to as \textit{pre-design}). We also include pruning-based baselines AgentPrune~\cite{Zhang2024AgentPrune} and AgentDropout~\cite{Wang2025AgentDropout}, which dynamically remove redundant links in multi-agent communication graphs (referred to as \textit{post-prune}). Since SafeSieve also belongs to the pruning paradigm, these two methods are further compared under prompt injection and heterogeneous-agent settings.

\subsection{Main Results}

\paragraph{\textbf{Takeaway 1: Carefully designed communication graphs outperform single-agent methods, with post-pruning paradigms comprehensively surpassing pre-design approaches.}} Multi-agent collaboration demonstrates significant performance improvements, with SafeSieve achieving 94.01\% average accuracy on DeepSeek-V3, substantially exceeding single-agent Vanilla (89.59\%) and CoT (90.58\%) baselines as shown in Table~\ref{tab:agentperformance}. Post-pruning methods universally outperform pre-design approaches, with AgentPrune (92.78\%), AgentDropout (92.62\%), and SafeSieve (94.01\%) all surpassing GPTSwarm (91.15\%) and G-Designer (92.00\%). SafeSieve achieves the best performance among post-pruning methods, reaching 94.01\% on DeepSeek-V3 and 87.61\% on GPT-4o-mini, establishing itself as the optimal solution across both model scales.

\paragraph{\textbf{Takeaway 2: Task-dependent performance improvements exhibit differentiated characteristics.}} Mathematical reasoning tasks (GSM8K, SVAMP) show stable improvements of 2-3 percentage points, with GSM8K improving from 94.68\% to 96.27\% (+1.59 points) and SVAMP from 93.67\% to 96.60\% (+2.93 points). Complex collaborative tasks demonstrate more significant gains, with MMLU improving by 4.42 points (87.97\%→92.39\%), HumanEval by 6.58 points (88.43\%→95.01\%), and AQuA by 7.31 points (84.58\%→91.89\%). This reflects SafeSieve's capability to identify and preserve critical heterogeneous communication paths through semantic similarity scoring.

\begin{figure*}[t]
\centering
\includegraphics[width=\textwidth]{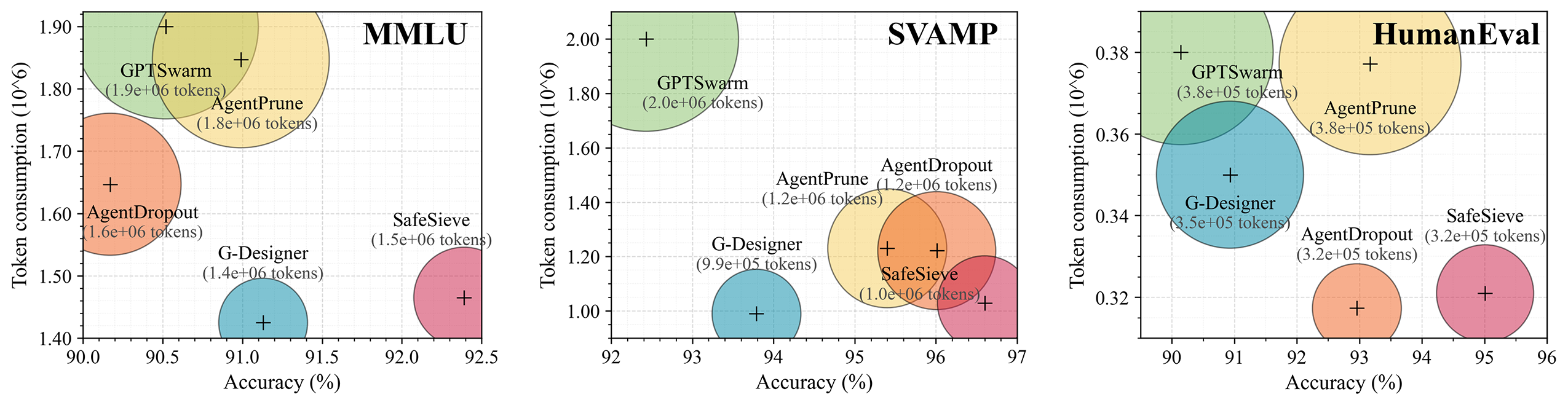}
\caption{Accuracy–efficiency trade-off across benchmarks. Each graph represents MAS method’s performance on one of three datasets: MMLU, SVAMP and HumanEval. It shows SafeSieve’s superior task-specific pruning capabilities.}
\label{fig:token_consume}
\end{figure*}

\paragraph{\textbf{Takeaway 3: Large and small models show differentiated improvements across task types, providing necessity for heterogeneous deployment.}} Large models (DeepSeek-V3) demonstrate greater improvements on complex tasks, with MMLU improving by 4.42 points and HumanEval by 6.58 points, showcasing advantages in handling complex collaboration. Small models (GPT-4o-mini) show higher improvements on structured tasks, with SVAMP achieving 5.7\% improvement (88.26\%→93.29\%) and GSM8K achieving 7.1\% relative improvement (87.45\%→93.61\%). This complementarity establishes the foundation for heterogeneous deployment, where large models excel at complex decision-making while small models are more efficient for structured tasks.

\paragraph{\textbf{Takeaway 4: SafeSieve achieves superior efficiency-accuracy trade-off positioning across all benchmarks.}} As demonstrated in Figure~\ref{fig:token_consume}, SafeSieve consistently occupies the optimal position in the accuracy-token consumption space. On MMLU, SafeSieve achieves 92.39\% accuracy with 1.47M tokens, outperforming GPTSwarm (90.52\%, 1.90M tokens) and AgentPrune (90.99\%, 1.85M tokens). On SVAMP, SafeSieve reaches 96.60\% accuracy with only 1.03M tokens compared to AgentPrune's 95.40\% at 1.23M tokens, achieving 16.3\% token reduction with 1.2 point accuracy improvement. For HumanEval, SafeSieve attains 95.01\% accuracy with 321K tokens versus AgentPrune's 93.17\% at 377K tokens. This efficiency advantage stems from SafeSieve's dual-stage scoring mechanism that precisely eliminates redundant paths while preserving critical collaborative links.

\paragraph{\textbf{Takeaway 5: Ablation experiments validate the effectiveness of three core components.}} As table~\ref{tab:ablation-safesieve} shows, historic feedback contributes significantly, with its removal causing accuracy to drop to 93.78\% (-1.23 points) while saving 30.0\% tokens. Heuristic initialization proves crucial, with its removal reducing accuracy to 94.41\% (-0.60 points) while saving 24.2\% tokens. The 0-extension clustering outperforms Top-k pruning, as replacing it with Top-k reduces accuracy to 93.13\% (-1.88 points), demonstrating the superiority of structure-aware clustering. The combination of all three components achieves optimal performance with 95.01\% accuracy and 27.8\% token savings, realizing the best balance between efficiency and performance.

\subsection{Robustness Analysis}

\begin{figure}[t]
\centering
\includegraphics[width=\columnwidth]{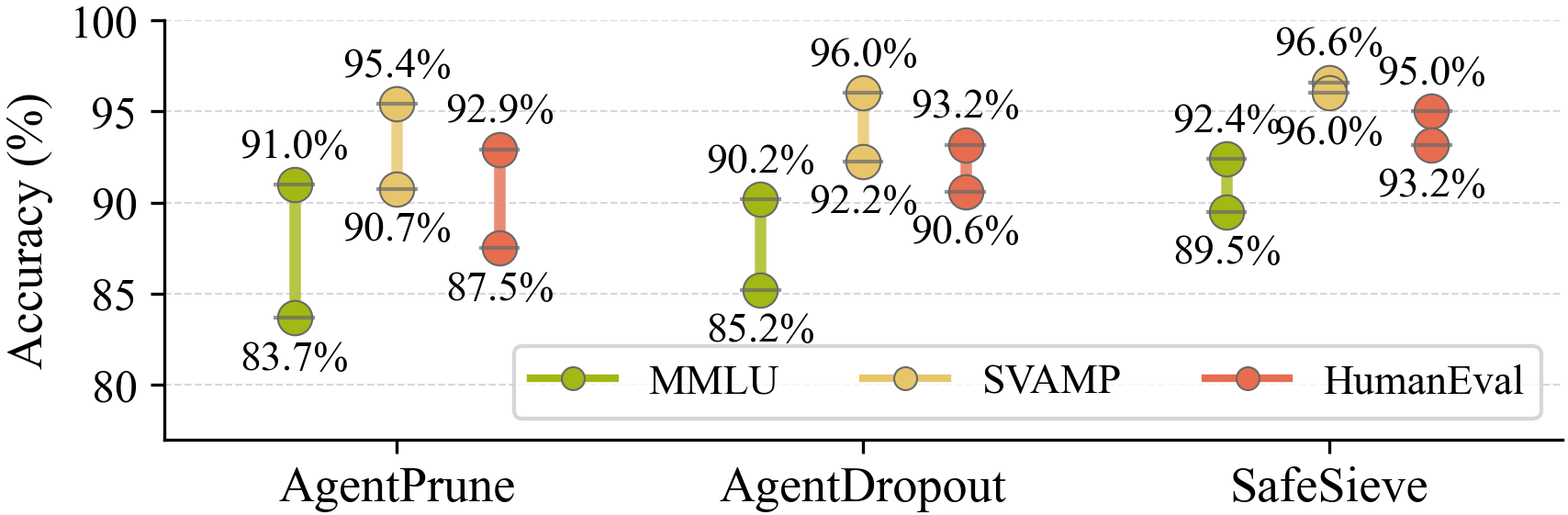} 
\caption{Accuracy drop of AgentPrune, AgentDropout, and SafeSieve when injecting low-quality agents into MMLU, SVAMP, and HumanEval tasks.}
\label{fig:injection}
\end{figure}

\paragraph{\textbf{Takeaway 1: Task characteristics determine vulnerability patterns to malicious agents.}} Knowledge-intensive MMLU proves most vulnerable, with SafeSieve accuracy dropping from 92.39\% to 89.50\% (-2.89 points), yet still outperforming AgentPrune's 7.32-point decline (90.99\%→83.67\%) as shown in Figure~\ref{fig:injection}. SVAMP mathematical reasoning shows minimal degradation, with SafeSieve declining only 0.56 points (96.60\%→96.04\%) while AgentPrune drops 4.66 points and AgentDropout drops 3.76 points. HumanEval programming tasks demonstrate moderate protection, with SafeSieve declining 1.84 points (95.01\%→93.17\%) compared to AgentPrune's 5.41-point drop and AgentDropout's 2.56-point drop.

\paragraph{\textbf{Takeaway 2: SafeSieve's triple defense mechanism ensures minimal degradation and superior robustness.}} Overall performance remains optimal with average drops of 1.23\% for SafeSieve, 2.21\% for AgentDropout, and 4.59\% for AgentPrune across three tasks, achieving a relative degradation rate of 1.33\% as detailed in Table~\ref{tab:intervention-drop}. Preventive defense assigns low weights (0.1-0.3) to suspicious agents compared to normal agents (0.7-0.9) through LLM semantic scoring. Responsive defense typically identifies malicious agents within 30 batches, automatically triggering pruning when cumulative scores fall below thresholds. Structural defense maintains network connectivity through 0-extension clustering, with accuracy fluctuations remaining under 3\% before and after pruning in MMLU experiments, avoiding \emph{information island} formation.

\input{tab/droprate}

\input{tab/hetecost}

\begin{figure*}[t]
\centering
\includegraphics[width=\textwidth]{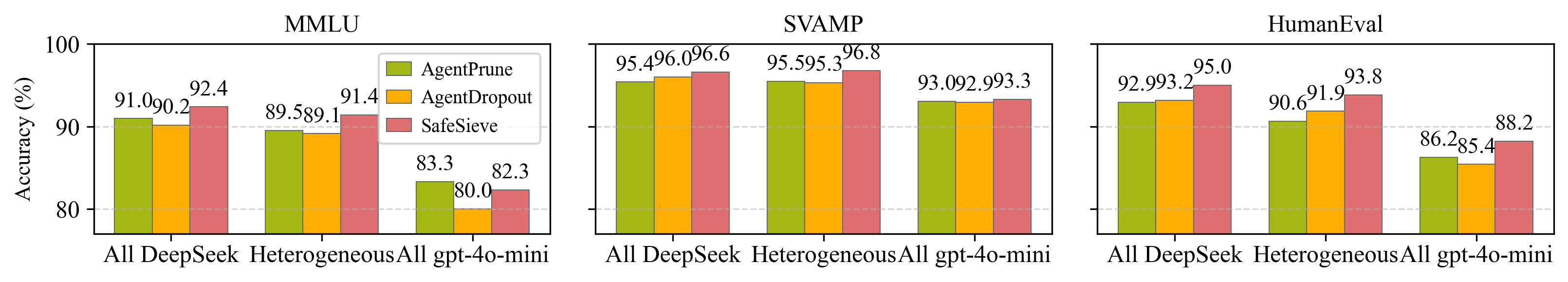} % Reduce the figure size so that it is slightly narrower than the column.
\caption{Performance and cost in heterogeneous settings. We compare AgentPrune, AgentDropout, and SafeSieve.}
\label{fig:hete}
\end{figure*}

\subsection{Heterogeneous Agent Collaboration}

\textbf{Experimental Setup Note.} DeepSeek-V3 serves as evaluation expert, chief commander, and answer extractor, while other subtasks are allocated to models including Qwen-72B, Kimi-K2, GPT-4o-mini, and LLaMA-8B.

\paragraph{\textbf{Takeaway 1: Large model \emph{commander} effect significantly reduces system costs.}} Total costs decrease by 13.3\% from AgentPrune's \textcent 115 to SafeSieve's \textcent 99.3 as demonstrated in Table~\ref{tab:hetecost}. Token allocation becomes intelligent, with DeepSeek-V3 consuming 995K tokens (40.7\% of total) for core reasoning and final answer extraction while small models handle 59.3\% of computational load. Cost optimization is significant, with DeepSeek-V3 costs reducing 15.4\% (\textcent 56.1→\textcent 47.5) and Kimi-K2 reducing 29.7\% (\textcent 36.7→\textcent 25.8). Small model usage increases but total costs decrease, with LLaMA-8B tokens increasing 23.1\% but costs rising only \textcent 0.01, and GPT-4o-mini tokens increasing 6.7\%. The \emph{1+4} collaboration mode achieves optimal cost-effectiveness ratio on SVAMP as shown in Figure~\ref{fig:hete}.

\paragraph{\textbf{Takeaway 2: Task-dependent heterogeneous effects exhibit \emph{barrel principle} characteristics.}} SVAMP mathematical reasoning shows heterogeneous advantages with 96.77\% accuracy, slightly exceeding homogeneous configuration's 96.60\% (+0.17 points) while reducing costs by 31.2\% (\textcent 28.2 → \textcent 19.4). MMLU knowledge tasks are limited by weaker models, with heterogeneous accuracy at 91.42\% falling below homogeneous 92.39\% (-0.97 points), validating the ``barrel effect" in knowledge-intensive tasks. HumanEval programming tasks maintain competitiveness with 93.78\% accuracy, declining only 1.23 points (vs. homogeneous 95.01\%) while reducing costs by 14.0\%.

\section{Conclusion}

We propose \textbf{SafeSieve}, a principled pruning framework for multi-agent collaboration that unifies semantic initialization with experience-guided refinement. It provides GPU-free sparsing strategy. Experiments across six benchmarks, including reasoning and coding tasks, show SafeSieve outperforms baselines with up to 6.58\% accuracy gains and 30\% token reduction. Furthermore, it demonstrates robustness against agent injection and excels in heterogeneous settings varying in model scale, validating the efficacy of structure-aware pruning for efficient LLM cooperation.

\bibliography{aaai2026}

\end{document}

%% file: tab/tab_agentperformance.tex
% Please add the following required packages to your document preamble:
% \usepackage{booktabs}
% \usepackage{graphicx}
% \usepackage[table,xcdraw]{xcolor}
% Beamer presentation requires \usepackage{colortbl} instead of \usepackage[table,xcdraw]{xcolor}

\begin{table*}[t]
    \centering
    \small
    \setlength{\tabcolsep}{4pt}
    \renewcommand{\arraystretch}{1.1}
    \begin{tabularx}{\textwidth}{l|>{\centering\arraybackslash}X|>{\centering\arraybackslash}X>{\centering\arraybackslash}X>{\centering\arraybackslash}X>{\centering\arraybackslash}X>{\centering\arraybackslash}X>{\centering\arraybackslash}X|>{\centering\arraybackslash}X}
        \toprule
        \textbf{Method} & \textbf{Paradigm} & \textbf{MMLU} & \textbf{GSM8K} & \textbf{SVAMP} & \textbf{HumanEval} & \textbf{AQuA} & \textbf{MATH-500} & \textbf{Avg.} \\
        \midrule
        \multicolumn{9}{c}{Base model: deepseek-V3-671B} \\
        \midrule
        \rowcolor[HTML]{EFEFEF} 
        Vanilla & single & 87.97\% & 94.68\% & 93.67\% & 88.43\% & 84.58\% & 88.20\% & 89.59\% \\
        CoT & single & 89.31\% & 95.15\% & 93.94\% & 89.26\% & 85.42\% & 90.41\% & 90.58\% \\
        \rowcolor[HTML]{EFEFEF} 
        GPTSwarm & pre-design & 90.52\% & 94.83\% & 92.43\% & 90.14\% & 88.40\% & 90.56\% & 91.15\% \\
        G-Designer & pre-design & 91.13\% & 95.47\% & 93.79\% & 90.93\% & 89.63\% & 91.02\% & 92.00\% \\
        \rowcolor[HTML]{EFEFEF} 
        AgentPrune & post-prune & 90.99\% & 95.30\% & 95.40\% & 92.91\% & 90.30\% & 91.76\% & 92.78\% \\
        AgentDropout & post-prune & 90.17\% & 95.16\% & 96.01\% & 93.16\% & 91.37\% & 89.82\% & 92.62\% \\
        \rowcolor[HTML]{EFEFEF} 
        SafeSieve (ours) & post-prune & \underline{\textbf{92.39\%}} & \underline{\textbf{96.27\%}} & \underline{\textbf{96.60\%}} & \underline{\textbf{95.01\%}} & \underline{\textbf{91.89\%}} & \underline{\textbf{91.90\%}} & \underline{\textbf{94.01\%}} \\
        \midrule
        \multicolumn{9}{c}{Base model: gpt-4o-mini ($\sim$8B)} \\
        \midrule
        \rowcolor[HTML]{EFEFEF} 
        Vanilla & single & 77.81\% & 87.45\% & 88.26\% & 87.08\% & 71.42\% & 70.14\% & 80.36\% \\
        CoT & single & 78.43\% & 87.10\% & 86.24\% & 88.13\% & 65.00\% & 77.18\% & 80.35\% \\
        \rowcolor[HTML]{EFEFEF} 
        GPTSwarm & pre-design & 82.80\% & 89.14\% & 87.02\% & \textbf{89.32\%} & 78.40\% & 80.70\% & 84.56\% \\
        G-Designer & pre-design & \textbf{87.20\%} & 93.97\% & 90.29\% & 87.50\% & 80.07\% & 81.76\% & 86.80\% \\
        \rowcolor[HTML]{EFEFEF} 
        AgentPrune & post-prune & 83.30\% & 93.58\% & 93.05\% & 86.25\% & 84.71\% & 82.81\% & 87.28\% \\
        AgentDropout & post-prune & 80.01\% & 93.25\% & 92.90\% & 85.41\% & 84.87\% & 81.97\% & 86.40\% \\
        \rowcolor[HTML]{EFEFEF} 
        SafeSieve (ours) & post-prune & \underline{82.32\%} & \underline{\textbf{93.61\%}} & \underline{\textbf{93.29\%}} & \underline{88.20\%} & \underline{\textbf{84.90\%}} & \underline{\textbf{83.33\%}} & \underline{\textbf{87.61\%}} \\
        \bottomrule
    \end{tabularx}
    \caption{Performance comparison between SafeSieve and baseline reasoning frameworks. Results for Vanilla and CoT under DeepSeek-V3 are adapted from AgentDropout~\cite{Wang2025AgentDropout} except MMLU and MATH-500; GPT-4o-mini results are taken from AGP~\cite{li2025adaptivegraph} except for MATH-500 and post-prune paradigm. Other results are evaluated by us under the same computing environment. All methods that start with a basic topology are based on \textbf{full-connected} graph.}
    \label{tab:agentperformance}
\end{table*}

%% file: tab/ablation.tex
% Please add the following required packages to your document preamble:
% \usepackage{booktabs}
% \usepackage{graphicx}
% \usepackage[table,xcdraw]{xcolor}
% Beamer presentation requires \usepackage{colortbl} instead of \usepackage[table,xcdraw]{xcolor}

\begin{table}[t]
\centering
\small
\setlength{\tabcolsep}{3pt}
\renewcommand{\arraystretch}{1.05}
\begin{tabular}{l|r|r|r|r}
\toprule
\textbf{Method} & \textbf{Acc.} & \textbf{Prom.} & \textbf{Comp.} & \textbf{↓Tokens} \\
\midrule
Full-connected (No Prune) & 95.50\% & 321K & 123K & --- \\
No-Heuristic Cluster Prune & 94.41\% & 233K & 104K & 24.2\% \\
No-Historic Cluster Prune & 93.78\% & 219K & 92K & 30.0\% \\
Combined + Top-$k$ Prune & 93.13\% & 207K & 107K & 29.3\% \\
\rowcolor[HTML]{EFEFEF}
SafeSieve (Ours) & 95.01\% & 223K & 98K & 27.8\% \\
\bottomrule
\end{tabular}
\caption{Ablation results on HumanEval. Compared to full-connected baseline, SafeSieve achieves comparable accuracy while reducing token usage by 27.8\%.}
\label{tab:ablation-safesieve}
\end{table}

%% file: tab/droprate.tex
% Please add the following required packages to your document preamble:
% \usepackage{booktabs}
% \usepackage{graphicx}
% \usepackage[table,xcdraw]{xcolor}
% Beamer presentation requires \usepackage{colortbl} instead of \usepackage[table,xcdraw]{xcolor}

\begin{table}[t]
\centering
\small
\setlength{\tabcolsep}{4pt}
\renewcommand{\arraystretch}{1.1}
\begin{tabular}{l|ccc|c|c}
\toprule
\textbf{Method} & \textbf{MMLU} & \textbf{SVAMP} & \textbf{H.E.} & \textbf{Avg.} & \textbf{Rate} \\
\midrule
\rowcolor[HTML]{EFEFEF} 
AgentPrune & ↓4.99 & ↓3.80 & ↓4.97 & ↓4.59 & ↓5.14\% \\
AgentDropout & ↓1.67 & ↓2.81 & ↓2.16 & ↓2.21 & ↓2.40\% \\
\rowcolor[HTML]{EFEFEF} 
SafeSieve & \textbf{↓1.19} & \textbf{↓1.60} & \textbf{↓0.91} & \textbf{↓1.23} & \textbf{↓1.33\%} \\

\bottomrule
\end{tabular}
\caption{Accuracy drop (\textminus) under malicious agent intervention. SafeSieve shows minimal average drop.}
\label{tab:intervention-drop}
\end{table}

%% file: tab/hetecost.tex
% Please add the following required packages to your document preamble:
% \usepackage{booktabs}
% \usepackage{graphicx}
% \usepackage[table,xcdraw]{xcolor}
% \usepackage{tabularx}
% Beamer presentation requires \usepackage{colortbl} instead of \usepackage[table,xcdraw]{xcolor}

\begin{table*}[t]
\centering
\footnotesize 
\setlength{\tabcolsep}{2.8pt}
\renewcommand{\arraystretch}{1.0}
\begin{tabular}{l|l|ccc|ccc|ccc|ccc}
\toprule
\multirow{2}{*}{\textbf{Method}} & \multirow{2}{*}{\textbf{Model}} 
& \multicolumn{3}{c|}{\textbf{MMLU}} 
& \multicolumn{3}{c|}{\textbf{SVAMP}} 
& \multicolumn{3}{c|}{\textbf{HumanEval}} 
& \multicolumn{3}{c}{\textbf{Overall}} \\
\cmidrule{3-14}
& & \textbf{Toks} & \textbf{Cost} & \textbf{$\Delta$} 
  & \textbf{Toks} & \textbf{Cost} & \textbf{$\Delta$} 
  & \textbf{Toks} & \textbf{Cost} & \textbf{$\Delta$} 
  & \textbf{Toks} & \textbf{Cost} & \textbf{$\Delta$} \\
\midrule
% --- AgentPrune---
\multirow{5}{*}{\shortstack[l]{\textbf{Agent}\\\textbf{Prune}}}
& DeepSeek-V3     & 491K & ¢23.46 & --   & 287K & ¢13.73 & --   & 397K & ¢18.94 & --   & 1,175K & ¢56.13& -- \\
& GPT-4o-mini     & 168K & ¢4.41 & --   & 84K  & ¢2.21 & --   & 86K  & ¢2.26 & --   & 338K & ¢8.87 & -- \\
& Llama-8B        & 131K & ¢1.31 & --   & 44K  & ¢0.44 & --   & 141K & ¢1.41 & --   & 316K & ¢3.15 & -- \\
& Qwen2.5-72B     & 171K & ¢6.19 & --   & 43K  & ¢1.56 & --   & 54K  & ¢1.97 & --   & 268K & ¢9.72 & -- \\
& Kimi-K2         & 229K & ¢16.88 & --   & 139K & ¢10.26 & --   & 130K & ¢9.57 & --   & 498K & ¢36.71 & -- \\
\cmidrule{2-14}
& \textbf{Total} & \textbf{1,190K} & \textbf{¢52.25} & -- 
                  & \textbf{597K} & \textbf{¢28.19} & -- 
                  & \textbf{808K} & \textbf{¢34.15} & -- 
                  & \textbf{2,595K} & \textbf{¢114.59} & -- \\
\midrule
% --- AgentDropout ---
\multirow{5}{*}{\shortstack[l]{\textbf{Agent}\\\textbf{Dropout}}}
& DeepSeek-V3     & 483K & ¢23.04 & -1.8\% & 233K & ¢11.12 & -19.0\% & 371K & ¢17.72 & -6.4\% & 1,087K & ¢51.88 & -7.6\% \\
& GPT-4o-mini     & 164K & ¢4.29 & -2.6\% & 85K  & ¢2.23 & +1.1\%  & 101K & ¢2.65 & +17.4\% & 350K & ¢9.17 & +3.4\% \\
& Llama-8B        & 122K & ¢1.22 & -6.4\% & 56K  & ¢0.56 & +28.9\% & 108K & ¢1.08 & -23.5\% & 286K & ¢2.86 & -9.2\% \\
& Qwen2.5-72B     & 203K & ¢7.36 & +18.9\% & 66K  & ¢2.39 & +53.1\% & 99K  & ¢3.58 & +81.5\% & 368K & ¢13.33 & +37.1\% \\
& Kimi-K2         & 201K & ¢14.85 & -12.0\% & 52K  & ¢3.81 & -62.9\% & 76K  & ¢5.63 & -41.2\% & 329K & ¢24.28 & -33.9\% \\
\cmidrule{2-14}
& \textbf{Total} & \textbf{1,173K} & \textbf{¢50.77} & \textbf{-2.8\%}
                  & \textbf{492K}  & \textbf{¢20.11} & \textbf{-28.6\%}
                  & \textbf{755K} & \textbf{¢30.65} & \textbf{-10.2\%}
                  & \textbf{2,420K} & \textbf{¢101.53} & \textbf{-11.4\%} \\
\midrule
% --- SafeSieve ---
\multirow{5}{*}{\shortstack[l]{\textbf{Safe}\\\textbf{Sieve}}}
& DeepSeek-V3     & 489K & ¢23.35 & -0.5\%  & 215K & ¢10.25 & -25.3\% & 291K & ¢13.89 & -26.7\% & 995K & ¢47.49 & -15.4\% \\
& GPT-4o-mini     & 133K & ¢3.49 & -20.9\% & 69K  & ¢1.82 & -17.3\% & 158K & ¢4.16 & +84.0\% & 360K & ¢9.47 & +6.7\% \\
& Llama-8B        & 230K & ¢2.3 & +75.6\% & 70K  & ¢0.70 & +60.7\% & 88K  & ¢0.88 & -37.3\% & 388K & ¢3.88 & +23.1\% \\
& Qwen2.5-72B     & 178K & ¢6.46 & +4.4\%  & 51K  & ¢1.83 & +17.2\% & 120K & ¢4.36 & +121.4\% & 349K & ¢12.65 & +30.1\% \\
& Kimi-K2         & 203K & ¢14.97 & -11.4\% & 65K  & ¢4.79 & -53.3\% & 82K  & ¢6.06 & -36.7\% & 350K & ¢25.82 & -29.7\% \\
\cmidrule{2-14}
& \textbf{Total} & \textbf{1,233K} & \textbf{¢50.56} & \textbf{-3.2\%}
                  & \textbf{470K}  & \textbf{¢19.40} & \textbf{-31.2\%}
                  & \textbf{740K} & \textbf{¢29.35} & \textbf{-14.0\%}
                  & \textbf{2,442K} & \textbf{¢99.31} & \textbf{-13.3\%} \\
\bottomrule
\end{tabular}
\caption{Comparison of token usage and cost across heterogeneous LLM models under three pruning paradigms. $\Delta$ indicates relative cost difference w.r.t. AgentPrune baseline.}
\label{tab:hetecost}
\end{table*}